\documentclass[aps,prl,reprint,showpacs,preprintnumbers,amsmath,amssymb,citeautoscript]{revtex4-1} 

\usepackage{graphicx}
\graphicspath{{figs/}{./}}
\usepackage{floatrow}
\usepackage[outercaption]{sidecap} 
\usepackage{color}
\newcommand\FigureFile[1] {#1.eps}
\DeclareGraphicsExtensions{pdf}

\usepackage{dcolumn}
\usepackage{bm}
\usepackage{color}
\usepackage{multirow}

\newcommand\eq[1]                              
{
\begin{align*}
#1
\end{align*}
}

\newcolumntype{L}[1]{>{\raggedright\let\newline\\\arraybackslash\hspace{0pt}}m{#1}}
\newcolumntype{C}[1]{>{\centering\let\newline\\\arraybackslash\hspace{0pt}}m{#1}}
\newcolumntype{R}[1]{>{\raggedleft\let\newline\\\arraybackslash\hspace{0pt}}m{#1}}
\newcolumntype{N}{@{}m{0pt}@{}}
\newcommand\eql[2] 
{
\begin{equation}\label{#1}
\begin{split}
#2
\end{split}
\end{equation}
}

\newcommand\eqll[3] 
{
\begin{equation*}\label{#1}
#3
\end{equation*}
}

\newcommand\eqsl[1]                            
{
\begin{align}
#1
\end{align}
}

\newcommand\eqssl[2]                      
{
\begin{subequations}\label{#1}
\begin{align}
#2
\end{align}
\end{subequations}
}

\definecolor{xmgrace-green4}{rgb}{0.0,0.55,0.0}
\definecolor{Green}{rgb}{0.2,0.96,0.2}
\definecolor{Remarks}{rgb}{1,0.3,0.3}
\definecolor{Extra}{rgb}{0.2,0.2,1}
\definecolor{Blue}{rgb}{0.2,0.3,1}
\definecolor{Black}{rgb}{0,0,0}

\newcommand\COMMENTED[1] {}

\begin{document}

\title{Suns-V$_\textrm{OC}$ characteristics of high performance kesterite solar cells}

\author{Oki Gunawan}
\thanks{Both authors contributed equally to this work}
\author{Tayfun Gokmen}
\thanks{Both authors contributed equally to this work}
\author{David B. Mitzi}
\affiliation{IBM T. J. Watson Research Center, PO Box 218, Yorktown Heights, NY 10598 USA}
\date{\today}

\begin{abstract}
Low open circuit voltage ($V_{OC}$) has been recognized as the number one problem in the current generation of Cu$_{2}$ZnSn(Se,S)$_{4}$ (CZTSSe) solar cells. 
We report high light intensity and low temperature Suns-$V_{OC}$ measurement in high performance CZTSSe devices. 
The Suns-$V_{OC}$ curves exhibit bending at high light intensity, which points to several prospective $V_{OC}$ limiting mechanisms that could impact the $V_{OC}$, even at 1 sun for lower performing samples.  
These V$_{OC}$ limiting mechanisms include low bulk conductivity (because of low hole density or low mobility), bulk or interface defects including tail states, and a non-ohmic back contact for low carrier density CZTSSe. 
The non-ohmic back contact problem can be detected by Suns-$V_{OC}$ measurements with different monochromatic illumination. 
These limiting factors may also contribute to an artificially lower $J_{SC}$-$V_{OC}$ diode ideality factor. 
\end{abstract}

\maketitle

\section{I. Introduction}
\label{sec:intro}
Kesterite Cu$_{2}$nSn(Se,S)$_{4}$ (CZTSSe) devices are emerging as a promising thin-film solar cell technology, given significantly improving power conversion efficiencies \cite{Todorov2010, Barkhouse2012, Wang2013} --currently as high as 12.6\%--and predominant use of more abundant and less toxic elements. 
This technology, if successfully developed to reach power conversion efficiency (PCE) beyond 18\%, has the potential to replace the existing thin film solar technologies, such as Cu(In,Ga)(S,Se)$_{2}$ (CIGSSe) and CdTe, which have issues with elemental abundance and toxicity.
Despite the promising recent progress in performance, open-circuit voltage, $V_{OC}$, remains as the number one problem in this technology \cite{Mitzi2013, Gunawan2010}.
Specifically, CZTSSe suffers from large $V_{OC}$ deficit i.e. the difference between the band gap $E_{g}$ and the open circuit voltage: $V_{OC,def} = E_{g}/q - V_{OC}$,  where $q$ is the electron charge.
The record CZTSSe device with 12.6\% power conversion efficiency (PCE) has $V_{OC} = 0.513$ V ($E_{g} = 1.13$ eV and $V_{OC,def} = 0.617$ V), which corresponds to only 57.8\% of the maximum $V_{OC}$ allowed by the Shockely-Quisser (SQ) limit \cite{Rau2004}.
In contrast, a record CIGSSe device with PCE of 20.8\% \cite{Jackson2014} has $V_{OC}$ of $0.757$ V ($E_{g} \sim 1.18$ eV and $V_{OC,def} = 0.423$ V), corresponding to 81.3\% of the maximum $V_{OC}$ according to the SQ limit.

Prior studies have discussed several possible factors contributing to the $V_{OC}$ deficit problem in CZTSSe such as interface recombination \cite{Gunawan2010}, low minority carrier lifetime \cite{Gunawan2010, Repins2013} and electrostatic potential fluctuations and tail states \cite{Gokmen2013}.
In this study we present another aspect of $V_{OC}$ limitation in CZTSSe, as revealed by high intensity and low temperature Suns-$V_{OC}$ measurements. 

\begin{figure}[bp]
\includegraphics[scale=0.93]{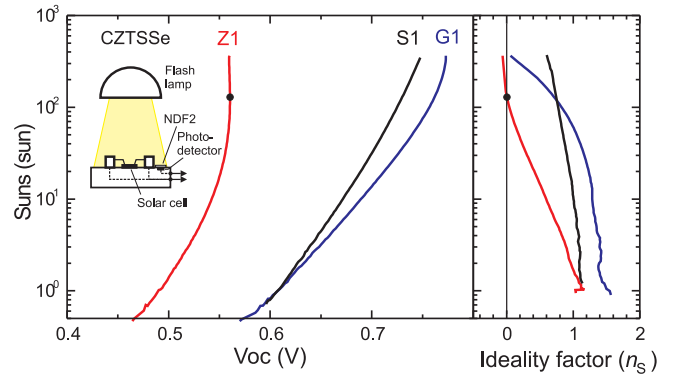}	
\caption{\label{fig:Fig01} High intensity Suns-$V_{OC}$ setup using a modified Sinton tool (inset) for a high performance CZTSSe ``Z1", CIGSSe ``G1" and mono-crystalline silicon ``S1" solar cell. 
\textit{Solid circle}: The Suns-$V_{OC}$ reversal point where the ideality factor drops to zero. \textit{Right panel}: The ideality factor $n_{S}$ as a function of light intensity.}
\end{figure}

\section{II. Experimental}
\label{sec:experimental}

\begin{table*}[tp]
\caption{\label{tbl:I}
Device parameters of solar cells used in this study under simulated AM1.5G illumination. Sample ZA, ZB and ZC are CZTSSe devices with varying Cu composition (and carrier density). $V_{OC,def}$ is the $V_{OC}$ deficit ($E_{g}/q Ð V_{OC}$), $p$ and $\mu$ are the Hall carrier density and mobility of the absorber layer measured on separate identical film.
}
\begin{tabular}{|C{0.5cm} |C{1cm}| C{1.3cm}| C{0.7cm}|C{0.7cm} | C{0.9cm}| C{1.4cm}| C{0.7cm}| C{1.2cm}| C{1.5cm}| C{1.6cm} |C{1.3cm}|N}
\hline
\multirow{2}{*}{No} & \multirow{2}{*}{Device}	& \multirow{2}{*}{Type}	&  $Eff$  & 	$FF$	 &$V_{OC}$	& $J_{SC}$	&$E_{g}$   &$V_{OC,def}$  &  \multirow{2}{*}{{\large $\frac{[\textrm{Cu}]}{[\textrm{Zn}]+[\textrm{Sn}]}$}}  & $p$& $\mu$ &\\ 
                             &                                               &                                             &       (\%)   &       (\%)    &    (mV)          & (mA/cm$^{2}$)  &    (eV)    &       mV            &                                 &      (/cm$^{3}$)      &  (cm$^{2}$/Vs) & \\ \hline
      1      &           G1     &    CIGSSe      &       $14.7$  &  $73.4$ & $600.2$    & $33.4$     &  $1.17$ & $569.8$ &	n.a.	& $\sim 2\times10^{15}$	& 3.7 & \\  [3.1pt]   \hline
      2      &	         S1	 &    Silicon	&        $14.1$ &	 $56.8$ &  $607.8$	 & $40.9$   &   $1.12$ & $512.2$ &	n.a.   &	    --	                            &  --   &\\ [3.1pt] \hline
     3	     &           Z1	&    CZTSSe	&      $11.7$  &	$67.3$ &	$494.0$	&  $35.2$	&     $1.13$ &	$636.0$	& $\sim0.82$	& $\sim2\times10^{15}$ &	 $0.45$ &\\ [3.1pt] \hline
     4	     &           Z2	 &   CZTSSe	&    $7.1$	 &    $49.1$   &	$459.6$	&   $31.6$ &	$1.13$	& $670.4$	 & $\sim0.8$	& --	& --&\\ [3.1pt] \hline 
    5	     &          ZA	 &   CZTSSe	&     $11.3$ &  $66.8$    & $481.4$	&  $35.2$	 &   $1.13$    &	$648.6$	& $0.82\pm0.4$ & $6.5\times10^{16}$ & $0.43$ &\\ [3.1pt] \hline
    6	    &          ZB      & 	CZTSSe	&     $11.5$  &	$70.8$  & $480.8$	& $33.7$	& $1.13$	& $649.2$	 & $0.86\pm0.4$ &	$1.6\times10^{17}$ &	$0.37$ &\\ [3.1pt] \hline
    7	    &          ZC     &       CZTSSe	&    $6.3$	  &    $51.5$  &  $416.4$	& $29.5$	&  $1.10$	& $683.6$	 &  $0.96\pm0.4$ &	$1.8\times10^{18}$ &	$0.56$ &\\ [3.1pt] \hline
\end{tabular}
\end{table*}

To perform a high intensity Suns-$V_{OC}$ measurement we utilize the Sinton Suns-$V_{OC}$ Illumination Voltage Tester with some modifications \cite{Sinton2000} as shown in the inset in Figure 1.
The basic system comes with a set of neutral density filters (NDF) at the outlet of the flash lamp. 
We remove these filters to increase the light intensity from $\sim1$ sun to $\sim300$ sun maximum. 
We then apply neutral density filters (NDF2) ($100\times$ attenuation) on top of the photodetector that monitors the light intensity to avoid saturation. 
Small probe testers are used to probe the two terminals of the solar cell to measure the $V_{OC}$. 
The flash light lasts for about $15$ ms and the light intensity and the $V_{OC}$ are recorded concurrently, as the light decays quickly from $\sim300$ sun to $0.3$ sun.
An example of the transient plot is shown in supplementary material (SM) Figure S1(b). 
The measurement can be repeated with a small shunt resistance ($R_{sh} = 1$  $\Omega$) across the device-under-test to measure the $J_{SC}$. 
Therefore, by combining Suns-$V_{OC}$ and Suns-$J_{SC}$  the $J_{SC}$-$V_{OC}$ plot can be obtained (see SM A and Fig. S2 for detail). 
Note that in most devices studied, especially at low light intensity ($< 1$ sun), $J_{SC}$ is proportional to the light (sun) intensity -- thus a description in terms of Suns-$V_{OC}$ or $J_{SC}$-$V_{OC}$ characteristics is identical and they are used interchangeably in this report. 
We also performed a wavelength-dependent Suns-$V_{OC}$ measurement by repeating the Suns-$V_{OC}$ curves with different color bandpass filters.  
This approach provides some depth sensitivity to the electrical measurements, as will be described later.

In the second part of the study, we developed a simple Sun-$V_{OC}$  (or $J_{SC}$-$V_{OC}$ ) measurement integrated to a standard existing solar simulator (Newport-Oriel, $1000$ W, $6"\times6"$ beam size) to facilitate immediate comparison of the light $J$-$V$ and $J_{SC}$-$V_{OC}$ curves (See SM B for details). 
A common method in performing Suns-$V_{OC}$  measurement is to use discrete neutral density filters to obtain different light attenuations (see. e.g. Ref. \cite{Krustok2010}); however this technique yields insufficient data resolution to accurately calculate the ideality factor. 
To solve this issue we developed a custom-made large area continuous neutral density (CND) filter, configured in a radial shape (see SM B) to achieve continuous light attenuation from $1$ to $\sim10^{-4}$ sun. 
This filter is made using a common overhead projector transparency printed with a radial grayscale pattern by ink-jet printing \footnote{O. Gunawan and B. Lei, Solar cell characterization system with an automated continuous neutral density filter, US patent publication 20120223733 A1 (2012).}.
The filter is driven by a stepper motor box to achieve a smooth, slow rotation that allows the $J_{SC}$  and $V_{OC}$  data to be recorded at varying intensities with fine resolution.

\section{III. Results}
\label{sec:results}

High intensity Suns-$V_{OC}$ measurements were performed on a high performance CZTSSe cell  (Z1), in comparison with analogous CIGSSe (G1) and silicon solar cells (S1) (Fig. 1). 
The detailed characteristics of these solar cells are presented in Table I. 
We can calculate the ideality factors from the Suns-$V_{OC}$ (or equivalently $J_{SC}$-$V_{OC}$) curves using the relationship $J_{SC} = J_{0}\exp(V_{OC}/n_{S}V_{T})$  and by assuming that the short circuit current $J_{SC}$ is proportional to the light intensity, i.e. $J_{SC} = SJ_{L1}$, where $S$ is the sun concentration factor (in  unit of ÒsunsÓ) and $J_{L1}$ is the photocurrent at $1$ sun. 
The ideality factor $n_{S}$ can be calculated as:
\eql{eq:IdealFac}
{
       n_{S} = [V_{T}d\ln J_{SC}/dV_{OC}]^{-1}=[V_{T}d\ln S/dV_{OC}]^{-1}
       \,,
}
where $V_{T} = k_{B}T/q$, $K_{B}$ is the Boltzmann constant and $T$ is the device temperature. 
The ideality factors of the Suns-$V_{OC}$ curves are shown in the right-hand side plot of Fig. 1.
The ideality factor tends to drop at higher light intensity for all cells, partly because of the Auger recombination effect that becomes more dominant at high carrier density \cite{Schmidt2000}.

We observe that the silicon (S1) and CIGSSe cells (G1) exhibit normal Suns-$V_{OC}$ curves that monotonically increase with higher light intensity.
In contrast, the CZTSSe cell Z1 shows a Suns-$V_{OC}$ curve that saturates at high sun intensity.
Furthermore, beyond a certain point (indicated by a solid circle) the Suns-$V_{OC}$ curve bends slightly backward. 
Correspondingly, the CZTSSe ideality factor derived from this curve ($n_{S}$) becomes anomalously low and turns negative at very high sun intensity (Fig. 1 right panel). 
Some CZTSSe samples show more severe backward bending as will be discussed in Fig. 6. 
This behavior suggests some mechanism that limits the $V_{OC}$. 
We have also repeated this measurement for CZTSSe with various band gaps (with low carrier density $< 10^{17}$ /cm$^{3}$) and observe similar bending behavior in all devices, as shown in SM Fig. S3. 

A similar observation can also be made at low light intensity Suns-$V_{OC}$ ($< 1$ sun) at low temperature from the temperature dependent study of the $J_{SC}$-$V_{OC}$ curves for the CIGSSe and CZTSSe cells as shown in Figure 2. 
We use $J_{SC}$-$V_{OC}$ measurement using the rotating CND filter as shown in Fig. 2(a) inset, this time focusing on the lower light intensity regime. 
The $J_{SC}$-$V_{OC}$ curves look normal (monotonic) for the CIGSSe cell at all temperatures and also normal for the CZTSSe at high temperature ($\sim300$ K). 
However, at low temperature ($< 140$ K) the CZTSSe $J_{SC}$-$V_{OC}$ curves exhibit bending behavior, very similar to what has been observed at ambient temperature under high sun intensity (Fig. 1).
 Fig 2(c) shows that at a constant 1 sun intensity, the $V_{OC}$ increases at lower $T$; however, as has been reported earlier \cite{Barkhouse2012, Gunawan2010}, due to the Suns-$V_{OC}$ bending behavior, the $V_{OC}$ drops at lowest temperatures for CZTSSe ($T < 150$ K).

\begin{figure*}[tp]
\includegraphics[scale=0.84]{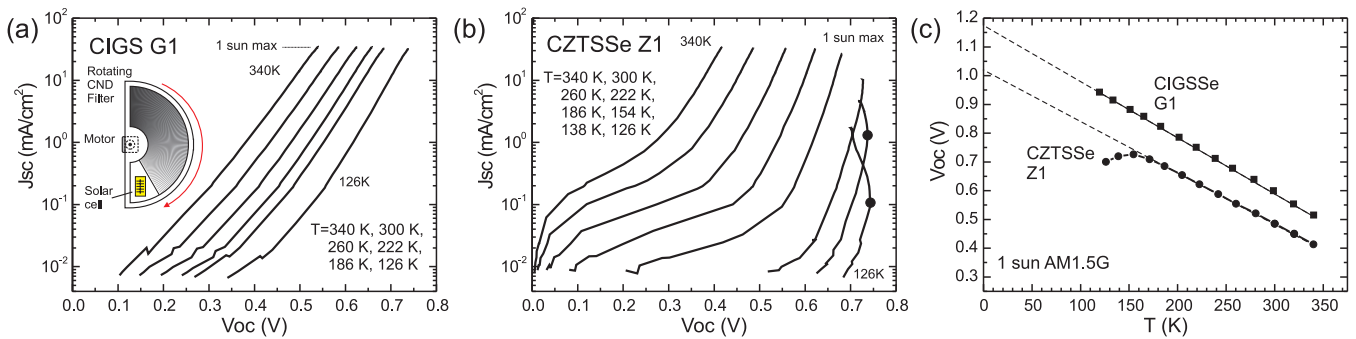}	
\caption{\label{fig:Fig02} Temperature dependence of the $J_{SC}$-$V_{OC}$ curves ($10^{-4}$ to $1$ sun) for: (a) CIGS; \textit{Inset}: The Continuous Neutral Density filter used to take the $J_{SC}$-$V_{OC}$ data with standard $1$ sun solar simulator; (b) CZTSSe;  and (c) $V_{OC}$ vs. temperature plot for both compounds. }
\end{figure*}

Another important set of information is obtained in a study of CZTSSe with varying [Cu]/([Zn]+[Sn]) ratio. 
Higher Cu-content leads to higher majority carrier (hole) density \cite{Nagaoka2013, Gunawan2014}. 
Fig. 3(a) shows that the high intensity $J_{SC}$-$V_{OC}$ bending only occurs for the sample with the lowest carrier density (ZA). 
Similarly, this behavior is also confirmed in low temperature measurement [Fig. 3(b)], i.e. only sample ZA shows a $V_{OC}$-$T$ curve that bends within the lowest temperature range ($T < 150$ K). 

\section{IV. Discussion}
\label{sec:discussion}

With respect to the possible origins of the observed Suns-$V_{OC}$ bending or pinning behavior in CZTSSe, we can differentiate two kinds of behaviors:  first is ``pinning" where the $V_{OC}$ gets saturated beyond some light intensity and second is ``bending" where the Suns-$V_{OC}$ curve bends backwards, as for the CZTSSe data in Fig. 1. 
Three factors can account for these pinning or bending behaviors as detailed below:

\subsection{A. Conductivity}

The first possible issue is low bulk conductivity, due to low carrier density or low majority carrier mobility. In order to give insight into this mechanism we perform device simulations using the wxAMPS program \cite{Liu2012,[{The high light intensity simulation is achieved by multiplying the input AM1.5G spectrum by the light intensity factor (e.g. $S=100$ sun)}]test1}. 
In Figure 4(a) we show the Suns-$V_{OC}$ simulation results for a baseline CIGS model \cite{Gloeckler2005} at three different hole mobility values. 
As the mobility value is reduced from $25$ to $1$ cm$^{2}$/Vs the Suns-$V_{OC}$ pinning behavior starts to develop; the effect becomes even more pronounced once the hole mobility is further dropped to $0.1$ cm$^{2}$/Vs. 


\begin{figure*}[tbp]
\floatbox[{\capbeside\thisfloatsetup{capbesideposition={right,top},capbesidewidth=3.6cm}}]{figure}[\FBwidth]
{\caption{(a) High intensity $J_{SC}$-$V_{OC}$ measurement (up to $320$ suns) for CZTSSe samples with increasing $[\textrm{Cu}]/([\textrm{Zn}]+[\textrm{Sn}])$ ratio or carrier density ($p$) (see inset). (b) $V_{OC}$ vs. temperature profile for the Cu-poor and Cu-rich devices.}\label{fig:Fig03}}
{\includegraphics[scale=0.98]{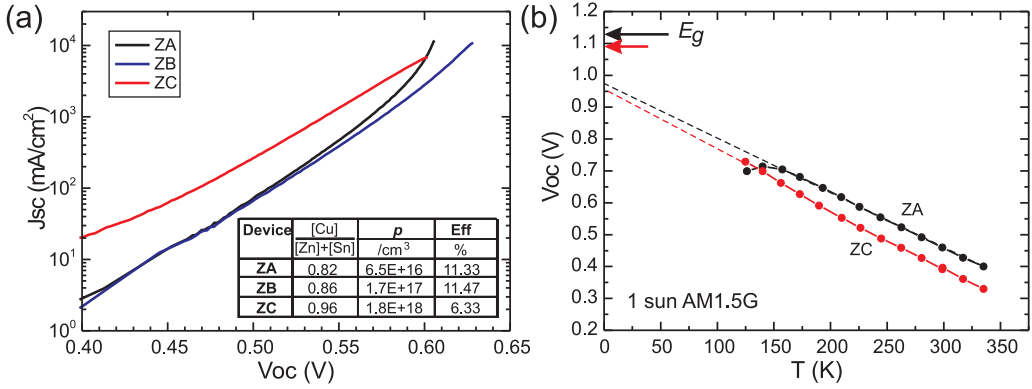}}
\end{figure*}

The physics of the observed behavior can be understood as follows. 
In a normal solar cell, the photo-generated electron-hole pairs in the depletion/junction region are separated by the built-in electric field (electrons are swept to the front and holes are swept to the back of the device) and therefore contribute to the $J_{SC}$. 
Because of the need for current continuity, in the short circuit condition this photo-generated current ($J_{SC}$) in the depletion region is maintained by majority carriers both at the front and the back of the device. 
At the back of the device majority carriers are holes and therefore it is the hole current, $J_{p}$, that can be considered to maintain the photo-generated current at any given Sun intensity. 
The hole current at the back can simply be written as  $J_{p} = qp\mu_{p}E$ where $q$ is the unit charge, $p$ is the hole (carrier) concentration, $\mu_{p}$ is the hole mobility and $E$ is the electric field at the back of the device. 
For the device with a hole mobility of $25$ cm$^{2}$/Vs, the necessary electric field at the back would be only $0.005$ V/$\mu$m, even at $100$ suns, and therefore this field does not noticeably affect the band bending throughout the device [Fig 4(b)]. 
However, for the device with a very low mobility (e.g., $0.1$ cm$^{2}$/Vs), the necessary electric field at the back is $1.25$ V/$\mu$m in order to maintained the $J_{SC}$ at $100$ Suns. 
Since this is a relatively large electric field, the band diagram completely changes throughout the device [Fig 4(c)] and the band bending mostly happens at the back of the device (giving rise to a voltage drop in the back region). 
The bending within the junction region is mostly screened due to the increase in hole concentration in the junction region because of the failure of the hole extraction from this region. 
Clearly this is an undesirable situation that reduces $J_{SC}$ compared to the ideal case and also results in the pinning of the $V_{OC}$ because of the voltage drop in the back region.  
Even under $V_{OC}$ condition [Fig. 4(c), right panel], there is still a significant band bending at the back that reduces the $V_{OC}$ due to significant hole current to cancel off the electron current.

Using this analysis we can also estimate the light intensity where the $V_{OC}$ pinning starts to occur. 
We set a maximum hole current ($J_{p}^{\max}$) that the back of the device can accommodate without substantially disturbing the band banding in the device, which we estimate at $E_{\max} \sim 0.1$ V/$\mu$m. 
Also, the short circuit current would be the product of the light intensity, $S$, and $J_{SC}$ at $1$ sun ($J_{L1}$). 
Therefore, using the equation  $J_{p} = qp\mu_{p}E$ and by equating these two currents, one could solve for the value of $S$ when the device starts to show a significant signature of $V_{OC}$ pinning: 
\eql{eq:S}
{
      J_{p}^{\max} = qp\mu_{p}E_{\max} = J_{L1}S
       \,.
}
We note that this simple equation is reasonably accurate to predict the Sun intensity where $V_{OC}$ starts to get pinned and the results are in good agreement with our simulation results shown in Figure 4(a).  
For example, in Fig. 4(a) we have $J_{L1} = 33$ mA/cm$^{2}$  for the baseline device ($\mu_{p} = 25$ cm$^{2}$/Vs). 
Using Eq. 2, we can estimate that the pinning should happen at approximately $S = 100$ and $10$ suns for the devices with $\mu_{p}= 1$ and $0.1$ cm$^{2}$/Vs respectively. 
Equation (2) also implies that the $V_{OC}$ pinning issue can be resolved (i.e., the bending pushed to higher Sun intensities) if the hole concentration and hole mobility product (bulk conductivity) is increased.

To assess the bulk conductivity, Van-der Pauw and Hall measurements were carried out on a set of high performance CZTSSe films, yielding carrier densities in the range of $10^{15}$ to $10^{18}$ /cm$^{3}$ (depending on [Cu]/([Zn]+[Sn]) content) while reference high performance CIGSSe films yield carrier density of $\sim2\times10^{15}$ /cm$^{3}$ as shown in Table I \footnote{The Hall measurements are performed on exfoliated CZTSSe and CIGS films. The same films produce a high performance CZTSSe (Eff $= 9-12$ \%) and CIGSSe devices (Eff $= 8-13$ \%). The films are exfoliated and transferred to a secondary glass substrate to isolate it from the underlying Mo metal layer that may otherwise shunt the Hall measurements.}.
 However, the CZTSSe mobility is notably lower, $\mu_{p}\sim (0.6 \pm 0.3)$ cm$^{2}$/Vs, compared to CIGSSe, $\mu_{p}\sim (4 \pm 0.5)$ cm$^{2}$/Vs. 
 For a baseline CZTSSe device with low carrier density $\sim10^{15}$ /cm$^{2}$ (e.g. sample Z1), this translates to lower bulk conductivity and as a result CZTSSe devices tend to suffer from higher series resistance and lower FF \cite{Barkhouse2012,Wang2013,Todorov2013}. 
 Thus, as further highlighted in the modeling results shown in Fig 4(a), it is reasonable that CZTSSe devices, especially those with low hole density ($<10^{17}$ /cm$^{2}$), more readily develop the Suns-$V_{OC}$ pinning due to its lower mobility.

 This $V_{OC}$ limiting mechanism is also consistent with the experimental data obtained from CZTSSe solar cells with increasing carrier density due to higher [Cu]/([Zn]+[Sn]) ratio. 
 Fig. 3 shows a set of CZTSSe devices with increasing [Cu]/([Zn]+[Sn])  ratio, which corresponds to higher carrier density $p$. 
 We observe that the Suns-$V_{OC}$ pinning disappears as $p$ increases with the mobility remaining roughly constant (Table I). 
 This behavior is also confirmed in low $T$ measurement, as shown in Fig. 3(b); the higher carrier density sample (ZC) does not show $V_{OC}$-$T$ pinning anymore at low temperature, at least down to $\sim120$ K. 
 Nevertheless, as our simulation in Fig. 4 shows, even for higher performance samples with a lower range of carrier density, this $V_{OC}$ limitation mechanism should be effectively benign at 1 sun (less than $10$ mV reduction), as long as carrier density and mobility remain above $\sim10^{16}$ /cm$^{3}$ and $\mu_{h} \sim 0.1$ cm$^{2}$/Vs, respectively (corresponding to a bulk conductivity of $0.016$ S/m).

 \subsection{B. Bulk or interface defects and tail states}
 
 Bulk and interface defects, including tail states introduce extra states in the band gap that could pin the Fermi level, thereby leading to Suns-$V_{OC}$ saturation. 
 In a solar cell, the $V_{OC}$ can be calculated from the separation of the electron ($E_{Fn}$ at the front contact) and hole ($E_{Fp}$ at the back contact) quasi Fermi levels: $V_{OC} = E_{Fn} - E_{Fp}$   [see Fig. 4(b)].  
 For a typical $p$-type solar absorber, the hole Fermi level is mainly determined by the free hole concentration and does not change with light illumination, as long as the excess carrier concentration does not exceed the free hole concentration. 
 In contrast, the position of the electron Fermi level, $E_{Fn}$, is determined by the balance between generation ($G$) and recombination ($R$) rate. 
 With light illumination, $E_{Fn}$ increases and stabilizes once the generation rate (which is proportional to light intensity) and recombination rate are equal. Most generally this condition can be represented by:
 \eql{eq:G}
{
      G(S) = R = \int_0^{\infty}\frac{1}{1+\exp[(E-E_{Fn})/k_{B}T]}\frac{g(E)}{\tau(E,n)}dE
       \,,
}
where $g(E)$ is the density of states in the conduction band including both extended and localized states (such as states that might arise from tails states or defect states) and $\tau(E,n)$ is the minority carrier lifetime that might depend on energy and excess carrier concentration.
In the simplified case, where $\tau(E,n)$   is just a constant, the above equation can be reduced to: 
$G=R= \int_0^{\infty}g(E)/(1+\exp[(E-E_{Fn})/k_{B}T])dE/\tau = \Delta n/\tau$.
In this simplified form and assuming density of states (DOS) for a clean semiconductor $(g(E) \propto \sqrt{E-E_{c}})$, one could arrive at the familiar relationship where $\Delta n \propto \exp(E_{Fn}/k_{B}T)$ and hence we obtain a $V_{OC}$ that increases monotonically with the sun intensity with ideality factor $n_{S} = 1$ (see detailed derivation in SM C).

 \begin{figure*}[tp]
\includegraphics[scale=0.895]{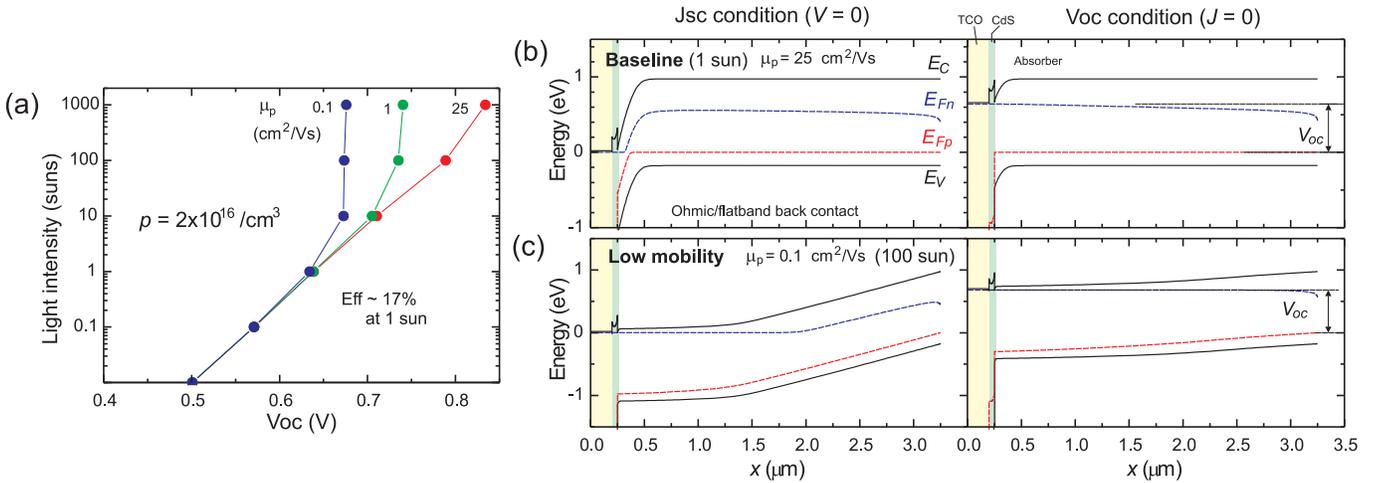}	
\caption{\label{fig:Fig04} (a) Suns-$V_{OC}$ simulation study of devices with varying hole mobility. (b), (c) Band diagrams at $J_{SC}$ (left column) and $V_{OC}$ (right column) for the (baseline) high mobility and low mobility device (at $100$ sun).}
\end{figure*}

  \begin{figure}[bp]
\includegraphics[scale=1.22]{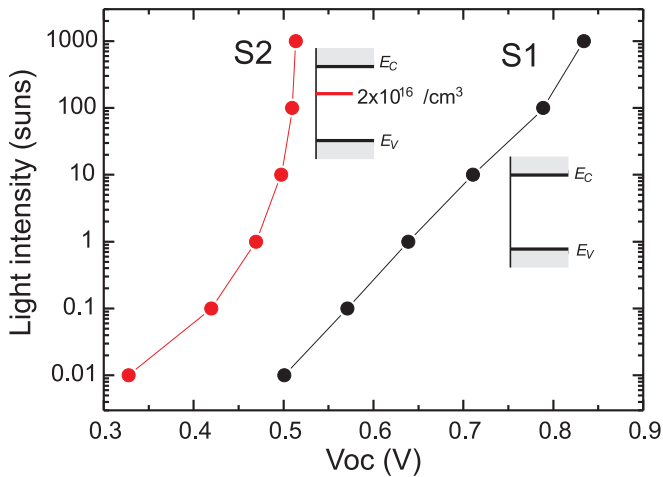}	
\caption{\label{fig:Fig05} Suns-$V_{OC}$ simulation of the impact of defect state in the absorber:  S1 (baseline device) and S2 (device with deep defect at $0.5$ eV below conduction band with density $2\times10^{16}$ /cm$^{3}$).}
\end{figure}

However for a disordered material, $g(E)$ will have states below the band gap (e.g. in the form of tail states \cite{Gokmen2013}). 
In addition, the minority carrier lifetime will become energy dependent, with lifetime being smaller for higher energy states since higher energy states are more delocalized. 
 Indeed, lifetimes that are factor of $10^{3}$ different have been observed for CZTSSe solar cells at low temperatures, as recombination changes from extended to localized states \cite{Gokmen2013,Gershon2013a,Gershon2013b}. 
 Interestingly, lifetime would also depend on the excess carrier concentration, where the shorter lifetimes would be observed for higher carrier concentrations. 
 This is also consistent with experimental observation of increasing lifetimes that is reported for CZTSSe solar cells as the time-resolved photoluminescence (TRPL) signal decays \cite{Wang2013,Todorov2013}. 
 The dependence of lifetime on carrier concentration can be understood in the context of electrostatic potential fluctuations \cite{Gershon2013a,Gershon2013b} or Auger recombination \cite{Quang1990}. 
 We note that the increased rate of Auger recombination in the presence of tail states and electrostatic potential fluctuations has been theoretically discussed in as Ref. \cite{Quang1990}. 
 Therefore, once all of these contributions are taken into account, the Suns-$V_{OC}$ pinning behavior that we observe in CZTSSe could also be due to the high density of bulk defects that result in band-tail states and electrostatic potential fluctuations. 
 Basically, since DOS increases with increasing energy and also higher energy states are more effective for recombination due to reduction in lifetime, $E_{Fn}$ (and hence $V_{OC}$) does not need to increase as much to satisfy the generation rate, thereby giving rise to $V_{OC}$ pinning behavior. 
 The same pinning behavior could also be the result of interface defects (arising from lattice mismatch and dangling bonds in heterojunction) as discussed by Ref. \cite{Turcu2003}. 
 In CIGSSe, the Fermi level pinning problem at the interface has been associated with interface defects such as donor-like anion (S,Se) vacancies \cite{Cahen1989} or with Cu/Cd  exchange \cite{Nakada1999}.

To illustrate this Suns-$V_{OC}$ pinning effect due to the presence of an electronic defect we perform a Suns-$V_{OC}$ simulation using the baseline CIGSSe model \cite{Gloeckler2005} (device ``S1") with parameters $ \eta = 16.7$ \%, $\textrm{FF}=79.4$ \%, $V_{OC} =0.639$ V, $J_{SC} = 33.1$ mA/cm$^{2}$. 
We create a ``defected device"  model  ``S2" by introducing a deep defect at $E_{C}$ - $0.5$ V, where $E_{C}$ is the conduction band edge, with a high defect density $N_{d} = 2\times10^{16}$ /cm$^{3}$. 
Device S2 yields solar cell parameters with notable reduction in $V_{OC}$ : $\eta=10.2$ \%, $\textrm{FF}=61.0$ \%, $V_{OC} =469$ V, $J_{SC} =35.6$ mA/cm$^{2}$. 
 The result of the simulation is shown in Fig. 5, with the ``defected device" S2 clearly exhibiting Suns-$V_{OC}$ pinning.

 \subsection{C. Non-ohmic back contact}

 \begin{figure*}[tbp]
\floatbox[{\capbeside\thisfloatsetup{capbesideposition={right,top},capbesidewidth=3.3cm}}]{figure}[\FBwidth]
{\caption{Device characteristics of a CZTSSe device Z2 with a suspected back contact problem: (a) $J$-$V$ characteristics; (b) Monochromatic $J_{SC}$-$V_{OC}$ test, using various band pass filters. The long wavelength light induces a more severe bending or lower $V_{OC}$ for the same $J_{SC}$.}\label{fig:Fig06}}
{\includegraphics[scale=0.95]{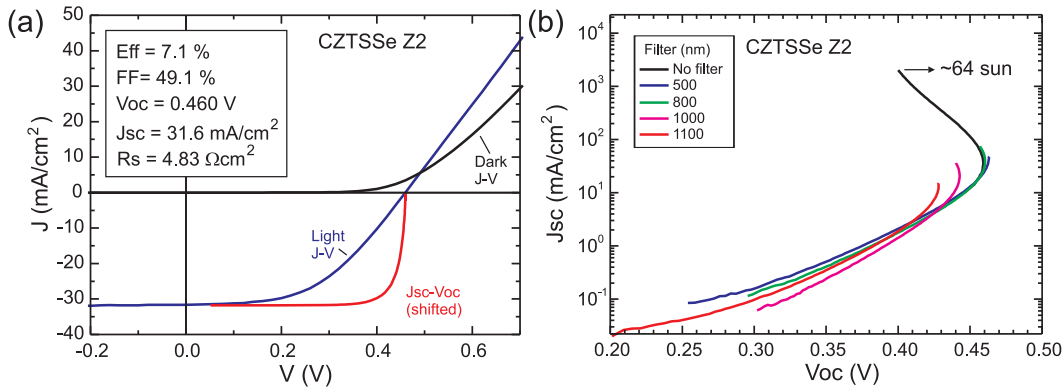}}
\end{figure*}

 While the two factors discussed above can explain the Suns-$V_{OC}$ pinning behavior, they cannot reproduce the bending behavior (i.e., measurements for which the $V_{OC}$ actually gets smaller with increasing light intensity) as found experimentally for some samples. 
 We consider two factors that can account for this feature. 
 One possible cause is a heating effect under high light intensity, as higher temperature will lead to lower $V_{OC}$ [e.g. see Fig. 2(c)]. The heating could be non-negligible at very high light intensity ($>100$ sun) given that all the absorbed light energy has to be converted to heat somewhere in the device for the open circuit condition. 
 Although heating may not be enough to give bending behavior alone, once combined with the pinning due to any of the two factors discussed above, it could contribute to mild backward bending behavior (and negative ideality factors). 
 However, the heating effect is not considered as the intrinsic $V_{OC}$ limiting factor and at $1$ sun this heating effect is expected to be negligible for the duration of the flash light  $< 10$ ms. 
 This is evident from the fact that in some samples [e.g. high carrier density sample ZC, Fig. 3(a)] the bending or pinning behavior is absent.

Nevertheless, in some mediocre CZTSSe devices, particularly those with very poor Fill Factor (FF), we observe severe backward bending behavior in the Suns-$V_{OC}$ curve Ð even at relatively low light intensity ($< 5$ sun). 
An example is shown for device CZTSSe Z2 in Fig. 6. 
In this kind of device, the cause of the backward bending may be attributed to a non-ohmic back contact. 
This issue has been well studied in silicon solar cells \footnote{S. W. Glunz, J. Nekarda, H. Mackel, and A. Cuevas, in 22nd European Photovoltaic Solar Energy Conference 2007} using Suns-$V_{OC}$ measurement of up to $1000$ suns and has also been suspected to be the problem in an earlier generation of CZTSSe \cite{Barkhouse2012, Wang2010}. 
Green \cite{Green1990} suggested a model consisting of a primary diode that represents the main photovoltaic (PV) junction and an opposing parasitic back contact (BC) diode in parallel with a back contact shunt resistance $R_{BC}$, as shown in Fig. 7(a) inset. 
This resistance is necessary for the whole device to operate at forward bias (otherwise an ideal back contact diode will block all forward bias current). When the back contact shunt resistance $R_{BC}$ is sufficiently high, appreciable photovoltage or $V_{OC}$ could develop across the BC junction producing an opposing voltage to the front junction. 
Physically, this back contact junction could arise at the CZTSSe/Mo(S,Se)$_{2}$ interface or Mo(S,Se)$_{2}$/Mo interface -- unfortunately we do not know the detailed electrical characteristics of the Mo(S,Se)$_{2}$ layer since it is buried deep within the absorber layer. 
However this model is sufficient to describe the Suns-$V_{OC}$ behavior qualitatively (Fig. 7).  
 

 Using this model we can attempt a fit to experimental data to gain more insight, as discussed in more detail in SM D.  
 Our device can be modeled as a standard solar cell junction ``PV" and a parasitic back contact junction ``BC" shunted by a resistance $R_{BC}$ as shown in Fig. 7(a). 
 The model has five independent parameters: $J_{L1A}$, $n_{A}$, $J_{L1B}/J_{0B}$, $n_{B}$ and $JR_{BC}$, where $J_{L1}$ is the 1 sun $J_{SC}$, $J_{0}$ is the dark reverse saturation current, $n$ is the ideality factor, and subscript A and B refers to the ``PV" and ``BC" diode, respectively.    
 Figure 7(a) presents the individual contributions of the PV and BC diode. 
 The main PV diode produces ideal increasing voltage with the light intensity and the BC diode produces small but increasing negative voltage that reduce the total $V_{OC}$. 
 Based on the parameters extracted, we can estimate the negative $V_{OC}$ contribution of the back contact at $1$ sun, which yields the relatively small value, $V_{OCB} = 13$ mV. 
 Note, however, that the model above does not take into account other factors that contribute to the Suns-$V_{OC}$ pinning or bending as described previously, such as a low conductivity effect, a bulk and interface defect effect and a prospective Auger recombination process (that dominates at very high intensity).  
 We also find that the dark reverse saturation current $J_{0B}$ of the BC diode (relative to the photocurrent $J_{L}$) is much larger than that of the primary diode ($J_{0A}$) indicating a very leaky junction. 
 This is expected for such a junction that is originally intended to serve as an ohmic contact. 
 Furthermore, the simulations [Fig. 7(b,c)] show that a lower reverse saturation current ($J_{0B}$) (more Schottky-like) and higher shunt resistance ($R_{BC}$) (less ohmic) back contact will lead to more severe Suns-$V_{OC}$ bending.  
 
    \begin{figure*}[tp]
\includegraphics[scale=1.0]{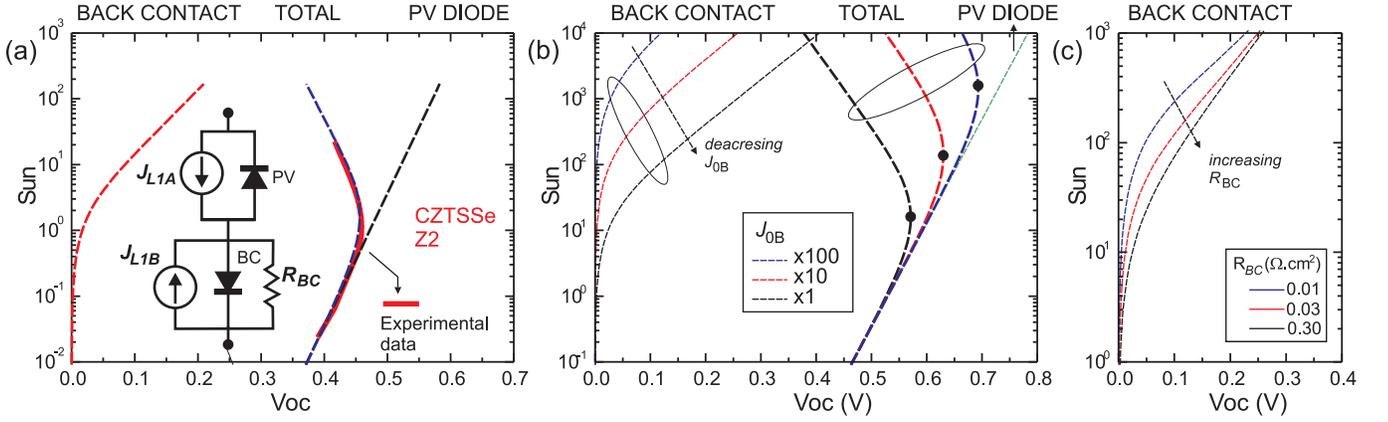}	
\caption{\label{fig:Fig07} (a) Circuit simulation to decompose the $V_{OC}$ contributions of the primary (PV) diode and the back contact (BC) diode from the experimental data. \textit{Inset}: The circuit model that includes a back contact diode and shunt resistance $R_{BC}$. (b) The effect of increasing BC dark current $J_{0B}$. \textit{Solid dot}: Suns-$V_{OC}$ reversal point where $n_{S} = 0$. (c) The effect of increasing $R_{BC}$. }
\end{figure*}

 Further evidence of the non-ohmic back contact problem can be obtained by performing Suns-$V_{OC}$ measurement with different color or band pass filters ($400$ to $1100$ nm) on the suspected device with very low FF (CZTSSe Z2, Fig. 6).  
 The low FF is mainly due to the large series resistance, as apparent from the significant difference between the light $J$-$V$ and shifted $J_{SC}$-$V_{OC}$ curve (also called pseudo $J$-$V$ curve) [Fig. 6(a)]. 
 Based on this back contact model we expect that, by shining a longer wavelength light, the photo-absorption at the back contact junction and thus the (negative) photovoltage will increase, thereby reducing the overall $V_{OC}$. 
 Indeed this is what we observe in Fig. 6(b).  
 The $J_{SC}$-$V_{OC}$ curves with longer wavelength illumination bend earlier and yield lower $V_{OC}$ for a given $J_{SC}$ value.  


 Finally, the back contact model can also provide a justification for the low temperature $J_{SC}$-$V_{OC}$ and $V_{OC}$  vs. $T$ behavior in Fig. 2. 
 We observe that the $J_{SC}$-$V_{OC}$ reversal points [solid dot in Fig. 2(b)] become visible below 1 sun and shift to lower light intensity at lower temperature ($T < 140$ K). 
  Regardless of the kind of junction (either $p$-$n$ or Schottky), lower temperature results in lower dark current $J_{0}$ and increase in the back contact photovoltage since $V_{OC} \sim \ln(J_{L}/J_{0})$  . 
  As a result, the $J_{SC}$-$V_{OC}$ reversal points occur at lower light intensity in the lower temperature $J_{SC}$-$V_{OC}$ curves. 
  This effect is simulated in Fig. 7(b), where lower dark current $J_{02}$ leads to more severe Suns-$V_{OC}$ bending. 
  The back contact issue could also contribute to the $J_{SC}$-$V_{OC}$  behavior of devices with different absorber conductivity, as discussed in Fig. 3. 
  Semiconductors with very low carrier density tend to yield worse ohmic contacts and thus should yield more significant Suns-$V_{OC}$  bending, as observed in the device with the lowest carrier density (ZA).


\subsection{D. Ideality factor difference}

The $V_{OC}$ reduction due to all different Suns-$V_{OC}$ pinning mechanisms can be estimated by drawing an asymptotic Sun-$V_{OC}$ line for the ideal ``PV diode," as shown in Fig. 7(a), and by noting the difference in $V_{OC}$ at $1$ sun between this line and the experimental data.
 Although the reduction in $V_{OC}$ at $1$ sun is negligible (only $\sim 3$ mV ) for a champion level CZTSSe solar cell [Fig. 1(a)], this reduction can be more severe ($\sim 13$ mV) for mediocre devices [as shown in Fig. 7(a)]. 
 The three factors discussed above are all likely present to varying degrees in our current portfolio of CZTSSe solar cells and therefore contribute to the Suns-$V_{OC}$ pinning and bending phenomena. 
 While it is difficult to decompose the contributions of each factor separately, we attempt to analyze this bending problem more quantitatively in order to investigate its impact on the $V_{OC}$ deficit at $1$ sun.

    \begin{figure}[bp]
\includegraphics[scale=1.25]{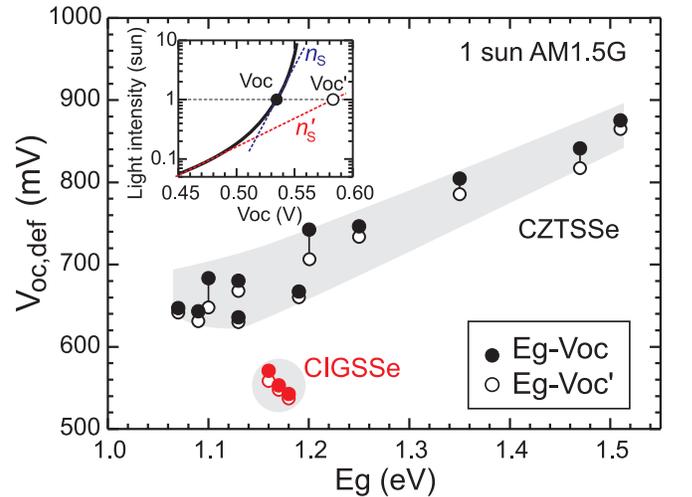}	
\caption{\label{fig:Fig08} $V_{OC}$ deficit in CZTSSe and CIGSSe as a function of bandgap (solid circles). At $1$ sun the effect of Suns-$V_{OC}$ pinning/bending on the $V_{OC}$ reduction (open circles) is relatively small, especially for the champion level device ($E_{g} \sim 1.13$ eV). \textbf{Inset}: Schematic illustration of a pinning/bending Suns-$V_{OC}$ curve. Solid (open) circle is the original $V_{OC}$ (estimated $V_{OC}^{'}$ if the pinning/bending effect is absence).}
\end{figure}

 \begin{figure*}[tp]
\floatbox[{\capbeside\thisfloatsetup{capbesideposition={right,top},capbesidewidth=3.6cm}}]{figure}[\FBwidth]
{\caption{(a) Light $J$-$V$ and ''pseudo $J$-$V$" curves (the $J_{SC}$-$V_{OC}$ curve offset by $J_{SC}$). \textit{Inset}: The efficiency and ideality factors extracted from LJV curves ($n_{L}$) and $J_{SC}$-$V_{OC}$ curves ($n_{S}$). (b)  $V_{OC}$ deficit vs. ideality factor difference $n_{L} - n_{S}$ (solid circles) and $n_{L} - n_{S}^{'}$ (hollow circles) (see text).}\label{fig:Fig09}}
{\includegraphics[scale=1.0]{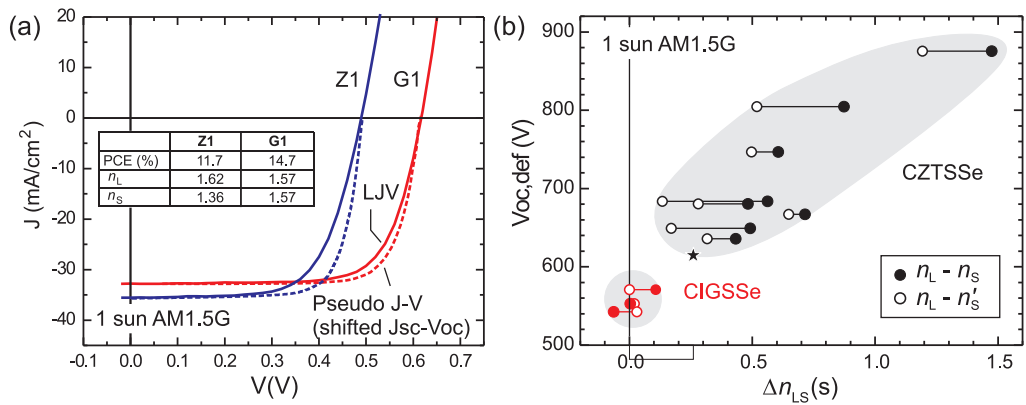}}
\end{figure*}

Figure 8 presents a plot of $V_{OC,def}$ of our CZTSSe devices as a function of bandgap. 
The CZTSSe samples show varying degrees of Suns-$V_{OC}$ pinning/bending behavior at $1$ sun.
 Using similar analysis to that presented in Fig. 7, we can estimate the expected $V_{OC}$ if the pinning behavior were absent (indicated as $V_{OC}^{'}$, which is higher than the original $V_{OC}$, thereby corresponding to lower $V_{OC}$ deficit as denoted by open circles in Fig. 8). 
 Evidently, at $1$ sun the Suns-$V_{OC}$ pinning/bending behavior has little impact on $V_{OC}$ deficit for the top performance samples ($E_{g} \sim 1.13$ eV). 
 Note that, in comparison with similar band gap CIGS devices and with the targeted $V_{OC,def}$ of $<500$ mV, we can conclude that the observed Suns-$V_{OC}$ bending/pinning effect does not account for the majority of the $V_{OC}$ deficit encountered in the CZTSSe-based devices (i.e., $>600$ mV).

One can also compare the ideality factor extracted from the Suns-$V_{OC}$ and the usual light $J$-$V$ curves. 
As shown in Fig. 1, the Suns-$V_{OC}$ bending behavior in CZTSSe artificially lowers the ideality factor $n_{S}$ extracted from this data. 
We can then define a new parameter $\Delta n_{LS}$, which is the difference of the ideality factors:
\eql{eq:Deltan}
{
      \Delta n_{LS} = n_{L} - n_{S}
       \,,
}
where $n_{S}$ is the Suns-$V_{OC}$ or $J_{SC}$-$V_{OC}$ ideality factor at $1$ sun, extracted from the asymptotic slope at $1$ sun (see SM C); and $n_{L}$ is the light $J$-$V$ ideality factor (at $1$ sun) derived from the following standard diode equation: $J = J_{0}\exp[(V-JR_{S})/n_{L}V_{T} - 1] + VG_{S} -J_{L}$, where $R_{S}$ is the series resistance, $G_{S}$ is the shunt conductance in the device and $J_{L}$ is the photogenerated current. 
We employ Sites' method to extract the four diode parameters---i.e., $J_{0}$, $n_{L}$, $R_{S}$ and $G_{S}$ as described in \cite{Sites1989} and \cite{Hegedus2004}.

An example of the ideality factor comparison of a CIGSSe and CZTSSe cell is presented in Fig. 9(a). 
We perform the Suns-$V_{OC}$ measurement using the rotating CND filter approach (SM B) in the same sitting right after light $J$-$V$ measurement (The $J_{SC}$-$V_{OC}$ curves are shifted down by $J_{SC}$ at $1$ sun for convenient comparison with light $J$-$V$ curves \cite{Sinton2000}). 
The ideality factor $n_{S}$ is extracted at the highest $J_{sC}$ point that corresponds to $1$ sun intensity using Eq. 1. 
In Fig. 9(a) we observe that the ideality factors $n_{L}$ and $n_{S}$ of the high performance CIGS cell are the same. 
In contrast, the CZTSSe Suns-$V_{OC}$ ideality factor ($n_{S}$) is smaller than the light $J$-$V$ ideality factor ($n_{L}$). 
We repeated this study in a collection of high performance CZTSSe and CIGSSe cells (with $\eta \sim 7 - 12.5$ \% and spanning the full range of CZTSSe bandgaps, $1.0 - 1.5$ eV) and investigated $V_{OC}$ deficit vs. bandgap and $\Delta n_{LS}$ as shown in Figures 8 and 9.  
We also present the data points of our recent 12.6 \% champion CZTSSe \cite{Wang2013} in Fig. 9(b) (shown as a star). 
As expected, it has nearly the lowest $V_{OC}$ deficit and the lowest $\Delta n_{LS}$ ($\sim0.25$), very close to the CIGSSe cluster. 
This suggests that the  $\Delta n_{LS}$ parameter can serve as another device quality indicator for thin film solar cells.

As discussed before, for mediocre devices the Suns-$V_{OC}$ bending can be significant even under $1$ sun condition and therefore this bending can artificially reduce $n_{s}$, giving rise to a large $\Delta n_{LS}$. 
However, using similar analysis to that presented in Fig. 8, we can define $n_{S}^{'}$, which is the Suns-$V_{OC}$ or J$J_{SC}$-$V_{OC}$ ideality factor at light intensities below $1$ sun where the pinning issue is absent. 
Interestingly, even after these corrections the difference $n_{L}$ - $n_{S}^{'}$ for the CZTSSe solar cell is still significant as illustrated in Figure 9(b). 
These arguments suggest that the difference between $n_{L}$ - $n_{S}^{'}$ derives from mechanisms beyond those discussed in this manuscript, therefore requiring further investigation.

\section{V. Summary}
In summary, we have presented high intensity and temperature dependent Suns-$V_{OC}$ (or $J_{SC}$-$V_{OC}$) measurements in a collection of high performance CZTSSe and CIGSSe cells. 
Unlike CIGSSe and silicon solar cells, many high performance CZTSSe cells (with typically low carrier density  and low [Cu]/([Zn]+[Sn]) exhibit Suns-$V_{OC}$ bending at high light intensity ($\sim100$ sun) at room temperature or even below $1$ sun at low temperature ($<140$ K).  
We discriminate two kinds of Suns-$V_{OC}$ behavior. 
The first is ``pinning" whereby the $V_{OC}$ gets saturated beyond some light intensity, which may be attributed to two factors: low bulk conductivity (mainly to low mobility) and the presence of bulk and interface defect states (including tail states) that could pin the Fermi level in CZTSSe. 
The second behavior is ``bending," where the Suns-$V_{OC}$ curve bends backwards at higher light intensity. 
This effect is attributed mainly to a non-ohmic back contact -- which is prevalent in CZTSSe with low carrier density (although heating effects do have the potential to contribute for intensities $>100$ suns). 
We have also demonstrated a technique to detect the non-ohmic back contact by performing Suns-$V_{OC}$ measurement employing different color band pass filters.  
 
The Suns-$V_{OC}$ pinning/bending symptom generally disappears for cells with higher carrier density due to increased bulk conductivity and better ohmic contact. 
However, these high carrier density samples ($p > 10^{17}$ /cm$^{3}$) have not historically been the devices with the highest performance.  
Also, the reduction in $V_{OC}$ at $1$ sun due to the pinning/bending behavior does not account for the majority of the observed large  $V_{OC}$ deficit for current-generation high-performance CZTSSe solar cells. 
 Therefore, even in cells where there is no Suns-$V_{OC}$ bending, there is still substantial $V_{OC}$ deficit. 
 We believe that this large $V_{OC}$ deficit is mostly accounted for by the band edge tail states in the CZTSSe material \cite{Gokmen2013}. 
 We also observe that the difference in the Suns-$V_{OC}$ and Light $J$-$V$ ideality factors ($\Delta n_{LS}$) grows with larger $V_{OC}$ deficit, suggesting that this parameter could serve as another indicator for device quality. 
 The techniques described here (high intensity Suns-$V_{OC}$ measurement, non-ohmic back contact detection and ideality factors comparison) are currently employed in our development of high performance CZTSSe devices to monitor and mitigate the $V_{OC}$ deficit issues, and can also be applied to other emerging solar cell technologies. 

\section{Acknowledgement}
This material is based upon work supported by the U.S. Department of Energy under Award Number DE-EE0006334. 
The work was conducted as part of a joint development project between Tokyo Ohka Kogyo Co., Ltd., Solar Frontier K. K. and IBM Corporation. 
The authors would like to thank Yiming Liu for his development of the wxAMPS simulation tool, Wei Wang and Teodor Todorov for supplying the high performance devices.

 \bibliography{CZTSSe_SunsVoc}

\end{document}